\documentclass[preprint]{aastex}

\slugcomment{to appear in the Astrophysical Journal Letters}
\shorttitle{Pigtail Molecular Cloud}
\shortauthors{Matsumura et al.}
\def\kms{\hbox{km s$^{-1}$}}
\def\VLSR{\hbox{$V_{\rm LSR}$}}

\def\sun{\hbox{$\odot$}}
\def\earth{\hbox{$\oplus$}}
\def\lesssim{\mathrel{\hbox{\rlap{\hbox{\lower4pt\hbox{$\sim$}}}\hbox{$<$}}}}
\def\gtrsim{\mathrel{\hbox{\rlap{\hbox{\lower4pt\hbox{$\sim$}}}\hbox{$>$}}}}

\def\arcdeg{\hbox{$^\circ$}}

\def\arcsec{\hbox{$^{\prime\prime}$}}

\begin{document}

\title{DISCOVERY OF THE PIGTAIL MOLECULAR CLOUD IN THE GALACTIC CENTER}

\author{Shinji Matsumura, Tomoharu Oka, Kunihiko Tanaka}  
\affil{Institute of Science and Technology, Keio University, 3-14-1 Hiyoshi, Yokohama, Kanagawa 223-8522, Japan.}
\author{Makoto Nagai}
\affil{Institute of Physics, University of Tsukuba, 1-1-1 Ten-nodai, Tsukuba, Ibaraki 305-8571, Japan.}
\author{Kazuhisa Kamegai}
\affil{Institute of Science and Technoogy, Tokyo University of Science, 2641 Yamazaki, Noda, Chiba, Japan.}
\and
\author{Tetsuo Hasegawa}
\affil{Joint ALMA Observatory, El Golf 40, Piso 18, Las Condes, Santiago, Chile.}


\begin{abstract}
This paper reports the discovery of a helical molecular cloud in the central molecular zone (CMZ) of our Galaxy.  This "pigtail" molecular cloud appears at $(l, b, \VLSR)\simeq (-0\arcdeg .7, +0\arcdeg.0, -70$ to $-30$ \kms), with a spatial size of $\sim 20\times 20$ pc$^2$ and a mass of $(2\mbox{--}6)\times 10^5$ $M_{\sun}$. This is the third helical gaseous nebula found in the Galactic center region to date. Line intensity ratios indicate that the pigtail molecular cloud has slightly higher temperature and/or density than the other normal clouds in the CMZ. We also found a high-velocity wing emission near the footpoint of this cloud. We propose a formation model of the pigtail molecular cloud. It might be associated with a magnetic tube that is twisted and coiled because of the interaction between clouds in the innermost $x_1$ orbit and ones in the outermost $x_2$ orbit.
\end{abstract}

\section{Introduction}
A region inside the $\sim 200$ pc radius of our Galaxy, called the "central molecular zone (CMZ)," shows a variety of phenomena that are unique to this environment. Molecular clouds in the CMZ are characterized by high temperature, high density, and large velocity dispersion, which are quite different from those of clouds anywhere else in the Galaxy (Morris \& Serabyn 1996). The interstellar magnetic field in the CMZ is suggested to be strong, although the field strength and geometry are still disputed (e.g., Ferri\`ere 2009).  A number of nonthermal linear filaments are also unique to this region. Their rigidity implies that a vertical magnetic field of $B\sim$mG, whether it is pervasive or not, penetrates the CMZ (e.g., Yusef-Zadeh, Morris, \&\ Chance 1984; Morris \&\ Yusef-Zadeh 1985). A minimum-energy analysis of low-frequency radio emissions indicates that magnetic fields of scale sizes greater than $50$ pc would be as weak as $\sim 10\, \mu$G, and at most $\sim 100\,\mu$G (LaRosa et al. 2005). On the other hand, submillimeter polarimetric maps show that the magnetic field permeating the CMZ is for the most part parallel to the Galactic plane (Novak et al. 2003).  

Thus far, two helical nebulae that may be related to the magnetic field have been found. The double helix nebula (DHN) has been found in the mid-infrared image at $\sim 100$ pc above the Galactic nucleus (Morris, Uchida, \&\ Do 2006).  It is thought to be a gaseous nebula associated with a torsional Alfv\'en wave propagating vertically away from the Galactic disk driven by rotation of the magnetized circumnuclear disk (CND). A map of the linear polarization at 10 GHz shows that the DHN has a highly ordered magnetic field along the ropes of the nebula (Tsuboi \&\ Handa 2010).  
Another helical nebula is the Galactic center molecular tornado (GCT) found in the CO {\it J} = 1--0 data set (Sofue 2007). The GCT is a helical-spur object of molecular gas at $\VLSR \sim 70$ \kms\ extending vertically from the Galactic plane at $l=+1.2\arcdeg$ to high latitudes of $b\sim \pm0\arcdeg .6$. It is thought to be formed by a magnetic squeezing mechanism.  
 
This paper reports the discovery of another helical gaseous nebula at $(l, b)\simeq (-0\arcdeg .7, +0\arcdeg.0)$. This was first noticed in the previously published CO {\it J} = 1--0 data to have a beautiful helix shape and an angular size of $\sim 0\arcdeg .15\times 0\arcdeg .15$ (Fig.1; see also Fig.4, $\VLSR=-50$ to $-40$ \kms\ of Oka et al. 1998).  It was named the "pigtail molecular cloud" (hereafter referred to as "the pigtail"). This pigtail appears to arise from a large molecular cloud at $\VLSR \sim -40$ \kms (hereafter referred to as "the $-40$ \kms\ cloud"), being located near the Sgr C HII region and the bottom of the western radio lobe (Handa et al. 1987). The pigtail is the most beautiful helical gaseous nebula found in the central environment of our Galaxy. This cloud is also apparent in the CO {\it J} = 3--2 data obtained using the Atacama Submillimeter Telescope Experiment (ASTE; Oka et al. 2007,2012) and in several millimeter wavelength molecular lines (presented in this paper). A possible origin of the pigtail is discussed in terms of a coiled magnetic tube scenario. 
\begin{figure*}[h]
\epsscale{1}
\plotone{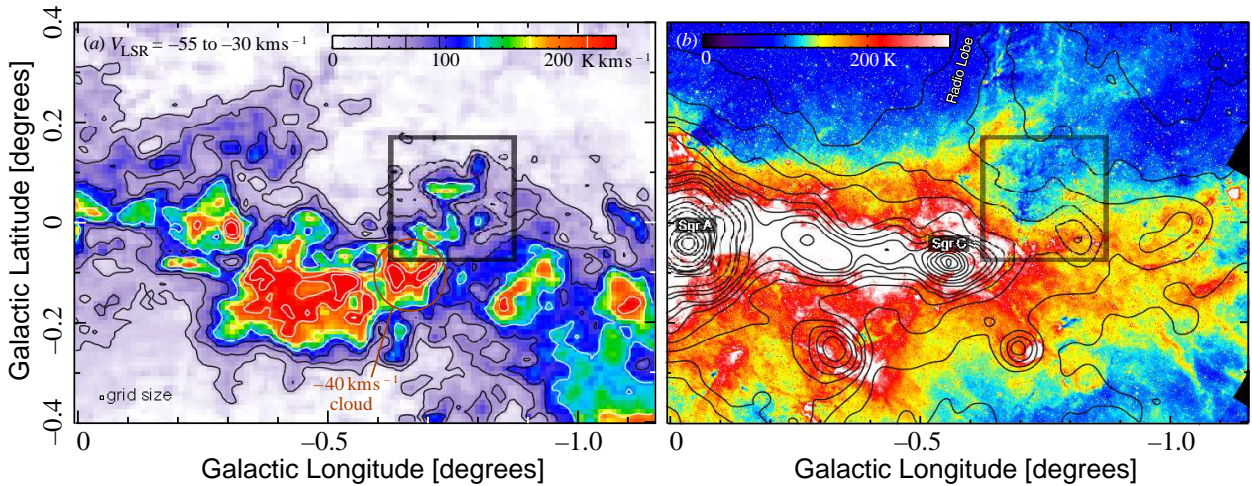}
\caption{({\it a}) Map of the velocity-integrated CO {\it J} = 1--0 emission. The velocity range for the integration is from $\VLSR=-55$ \kms\ to $-30$ \kms. The contours are 35, 65, 95, 125, 175, 225, and 275 K \kms\, and the white contour begins at 125 K \kms. The thick rectangle indicates the area where the follow-up observations are performed. ({\it b}) 24 $\mu$m image taken using the Spizter Space Telescope (Carey et al. 2008). The countours show the 10 GHz radio continuum map (Handa et al. 1987). The thick rectange is the same as that in the panel ({\it a}).  \label{fig1}}
\end{figure*}
 
\begin{figure*}[h]
\epsscale{1}
\plotone{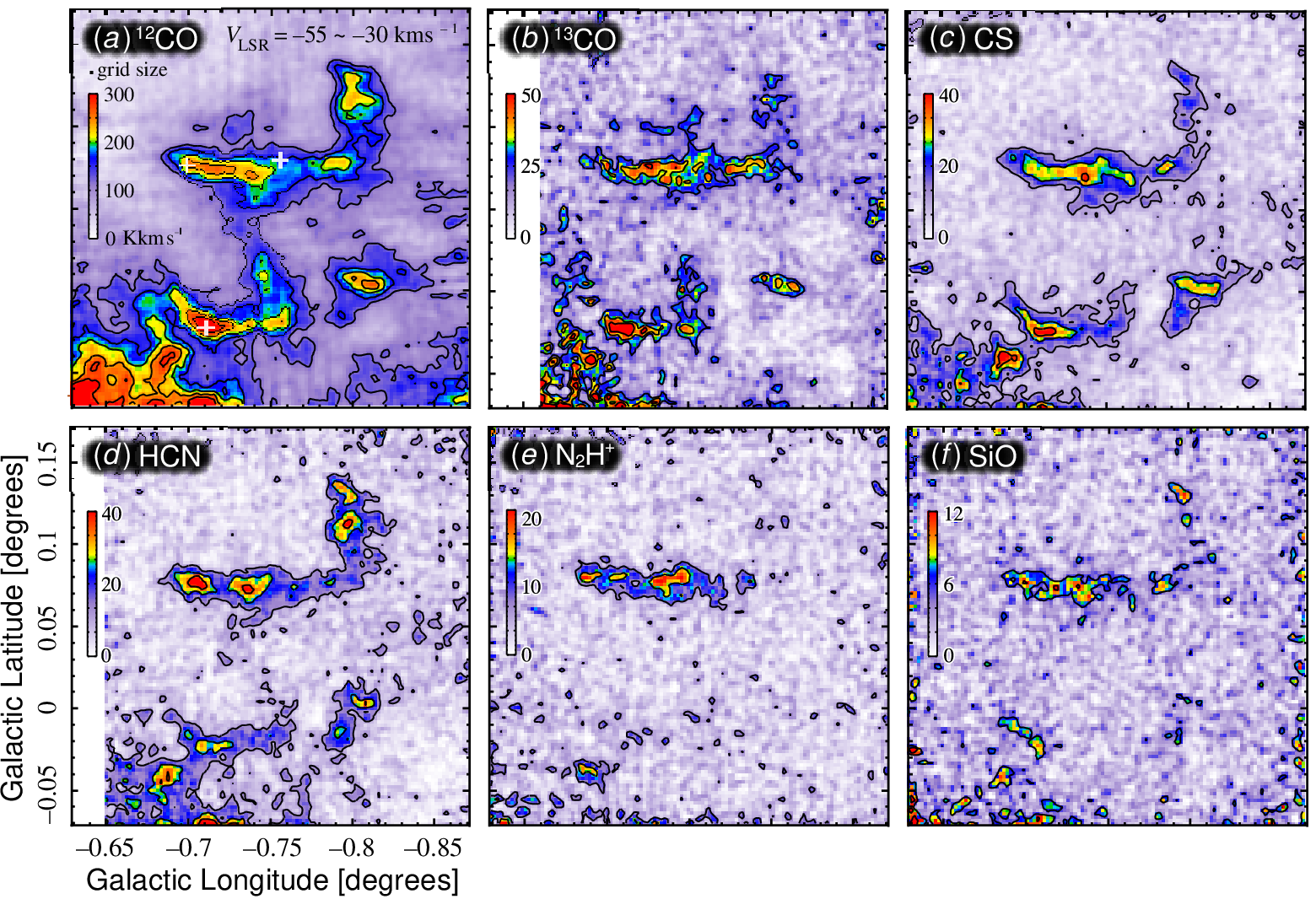}
\caption{Maps of velocity-integrated molecular line emission. The velocity range for the integration is from $\VLSR=-55$ \kms\ to $-30$ \kms. ({\it a}) CO {\it J} = 1--0 line. The contours are set at 40 K \kms\ intervals from 120 K \kms. ({\it b}) $^{13}$CO {\it J} = 1--0 line. The contours are set at 12 K \kms\ intervals from 24 K \kms. ({\it c}) CS {\it J} = 2--1 line. The contour interval is 12 K \kms. ({\it d}) HCN {\it J} = 1--0 line. The contour interval is 12 K \kms. ({\it e}) N$_2$H$^+$ {\it J} = 1--0 line. The contour interval is 7 K \kms. ({\it f}) SiO {\it J} = 2--1 line. The contour interval is 6 K \kms.  \label{fig2}}
\end{figure*}

\begin{deluxetable}{lrccccc}
\tabletypesize{\scriptsize}
\tablecaption{Parameters of Observed Lines\label{tbl-1}}
\tablewidth{0pt}
\tablehead{
\colhead{Line} & \colhead{$\nu$ [GHz]} & \colhead{$n_{\rm crit}$ [cm$^{-3}$]} & \colhead{$\eta_{\mathrm{MB}}$} & \colhead{HPBW\tablenotemark{a}} & \colhead{$\Delta T_\mathrm{MB}$ [K]\tablenotemark{b}}
}
\startdata
CO \textit{J}=1--0 & 115.271204 & 400 & 0.31(3) & $15.2\pm 0.2\arcsec$ & 0.71\\
$^{13}$CO \textit{J}=1--0 & 110.201353 & $10^3$ & 0.43(3) & $15.2\pm 0.2\arcsec$  & 0.53\\
CS \textit{J}=2--1 & 97.980968 & $10^6$ & 0.45(3) & --- & 0.21\\
N$_{2}$H$^{+}$ \textit{J}=1--0 & 93.173809  & $2\times 10^5$ & 0.46(3) & --- & 0.23\\
HCN \textit{J}=1--0 & 88.631847 & $2\times 10^5$ & 0.47(2) & $18.6\pm 0.2\arcsec$ & 0.3\\
SiO \textit{J}=2--1 & 86.846998 & $10^6$ & 0.47(2) & $18.6\pm 0.2\arcsec$ & 0.18
\enddata

				   
\tablecomments{Values in parentheses show one standard deviation of uncertainty in the last digit.}
\tablenotetext{a}{Half-power beam widths at 112 GHz (CO and $^{13}$CO) and 89 GHz (HCN and SiO) are listed.  Errors are the standard deviations from the averages for the 25 beams of BEARS.}
\tablenotetext{b}{Noise levels are in 1$\sigma$.  }
\end{deluxetable}

%

\section{Observations}
Follow-up molecular line observations were carried out using the Nobeyama Radio Observatory 45 m radio telescope between February and March 2009. The observed lines are listed in Table 1 along with the frequency, critical H$_2$ density, beam efficiency, beamsize, and rms noise of the final maps.  All the observations were made with the focal-plane array SIS receiver, BEARS (Sunada et al. 2000), in the on-the-fly (OTF) mapping mode (Sawada et al. 2008).  The reference position was $(l,b)=(+0\arcdeg ,-1\arcdeg)$.  The antenna temperature was obtained by the standard chopper-wheel technique (Kutner \&\ Ulich 1981).  The telescope pointing was corrected every 1.5 h by observing the SiO maser source VX Sgr and therefore maintained to $\leq 3\arcsec $.  25 digital autocorrelators were used as spectrometers in the wide-band mode, which has a 512 MHz coverage (1700 \kms\  at 90 GHz) and a 500 kHz resolution (1.7 \kms). 

The obtained data were reduced using the NOSTAR reduction package.  The 25 beams of BEARS show variations of approximately $10\%$ in both beam efficiency and sideband ratio.  We calibrated the intensity scale of each beam by comparing it with that taken with the single-sideband (SSB) receiver S100 to obtain an accurate antenna temperature, $T^{*}_{\rm A}$.  We scaled the antenna temperature by multiplying it by $1/\eta_{\rm MB}$ to obtain the main-beam temperature, $T_{\mathrm{MB}}$.  The typical double-sideband (DSB) system noise temperature was 200--400 K.  The data sets cover an $\sim 0\arcdeg.2\times0\arcdeg.2$ area centered at $(l,b)=(-0\arcdeg .75, +0\arcdeg .05)$, covering the full extent of the pigtail.  All the data were resampled onto a $7\arcsec .5\times 7\arcsec .5 \times 1$ \kms\ grid to obtain the final maps.

\section{Results}
\subsection{Spatial Distributions}
Figure 2 shows velocity-integrated molecular line maps of the pigtail region.  In the $^{12}$CO map (Fig.2{\it a}), the pigtail exhibits a distinct helical appearance with a $\sim 15$ pc pitch.  At least two rounds of a conical helix can be traced in the $^{12}$CO map.  The most intense $^{12}$CO ($^{13}$CO and CS) emission is found at $(l,b)\simeq (-0\arcdeg .71, -0\arcdeg .02)$ in the first round (from the bottom) of the pigtail.  This position corresponds to that of the negative high-velocity wing that is described later in this subsection.  The northwestern head of the $-40$ \kms\ cloud is adjacent to the footpoint of the pigtail (see also Fig.1{\it a}).  We observe a clump at $(l, b)\simeq (-0\arcdeg .69, -0\arcdeg .04)$ that appears to correspond to the footpoint of the pigtail.  Another clump is found at $(l,b)\simeq (-0\arcdeg .8, 0\arcdeg .0)$, the relation of which to the pigtail is unclear.  

The two rounds of the pigtail are disconnected in the maps of the other five lines (Fig.2{\it b--f}).  We can barely trace the helical structure in the $^{13}$CO, CS, and HCN maps.  N$_2$H$^+$ and SiO maps show a straight cloud that corresponds to the second round and small clumps near the footpoint.  The various appearances in the molecular maps might be due to the significant variation in the column density, density, and/or molecular abundances along the pigtail.  The portions that disappeared in the CS and HCN lines might have density lower than $10^5$ cm$^{-3}$.  The appearances of N$_2$H$^+$ and SiO may be affected by the variation in molecular abundance.  The N$_2$H$^+$ line traces chemically matured dense molecular gas (e.g., Lee et al. 1996), whereas the SiO line traces shocked molecular gas (e.g., Ziurys, Friberg, \&\ Irvine 1989).  This paper does not go into details about the abundance variation in the pigtail.


\begin{figure*}[t]
\epsscale{.8}
\plotone{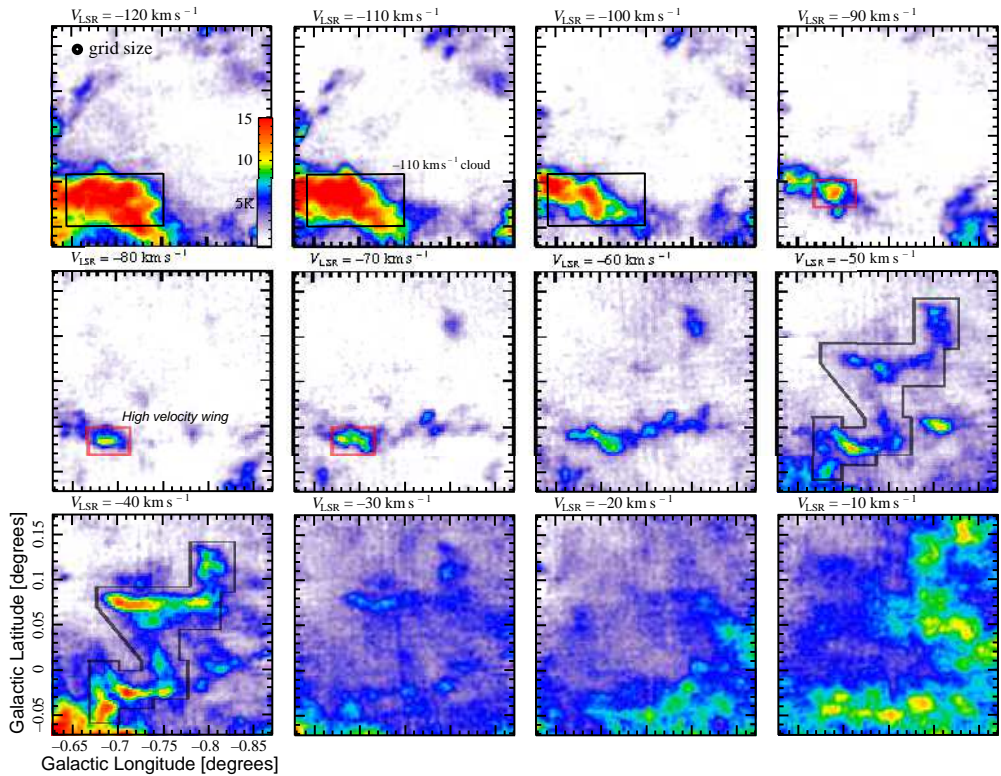}
\caption{Velocity channel maps of $^{12}$CO {\it J} = 1--0 emission.  The velocity width of each channel is 10 \kms, and the central velocity is indicated above each panel.  The thin solid line shows the integration area in the calculation of ratios for the pigtail, and the thick solid line shows that for the wing.  \label{fig3}}
\end{figure*}

\begin{figure}[t]
\epsscale{0.4}
\plotone{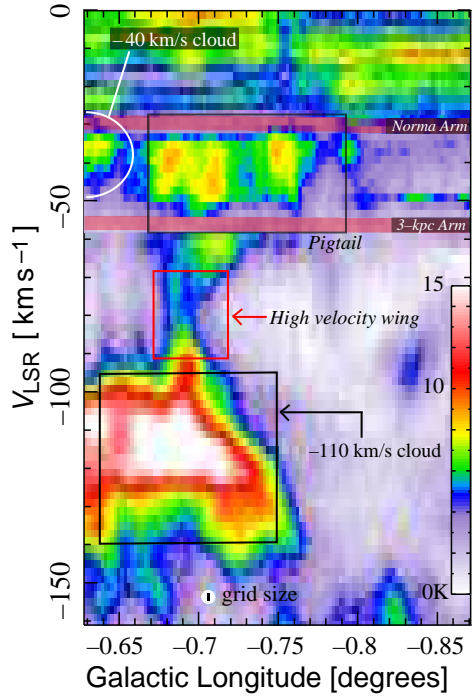}
\caption{Longitude-velocity map of CO {\it J} = 1--0 emission integrated over latitudes from $b=-0\arcdeg.03$ to $0\arcdeg.0$.   \label{fig4}}
\end{figure}

\begin{figure}[t]
\epsscale{1.0}
\plotone{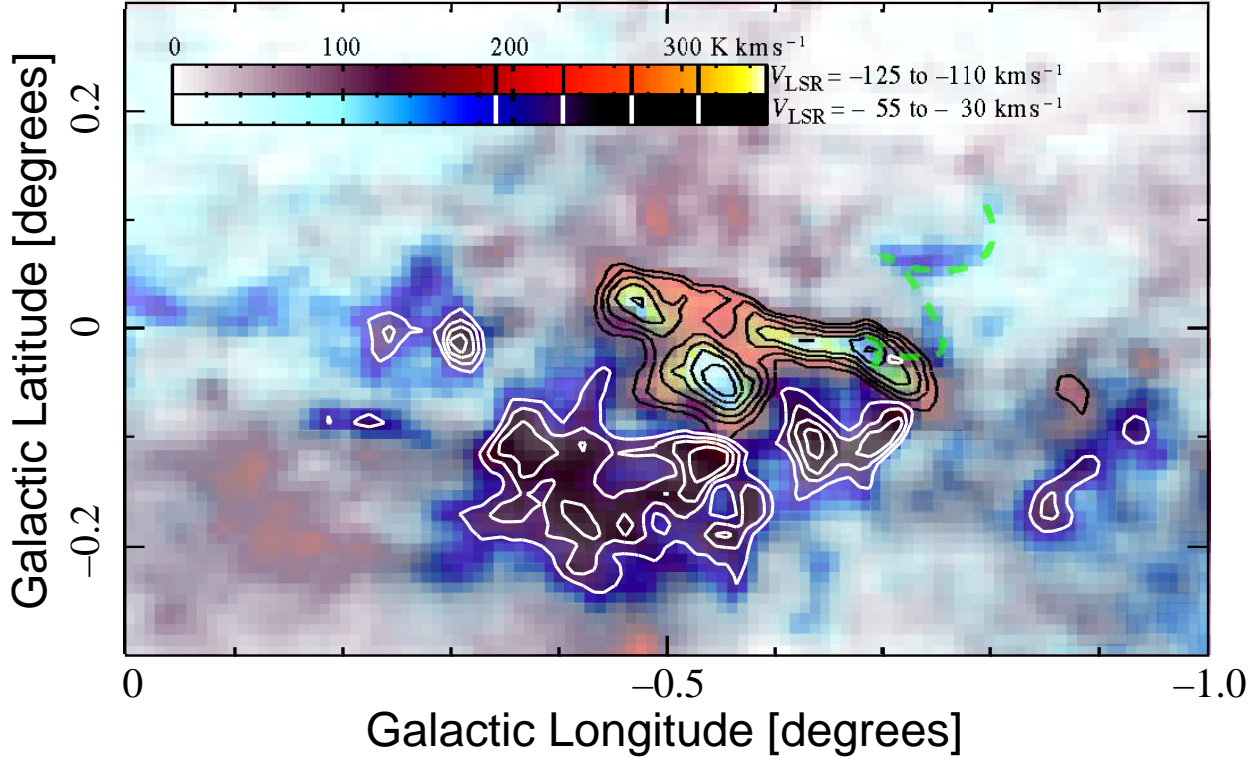}
\caption{Maps of $^{12}$CO integrated intensity for the velocity ranges from $-125$ to $-110$ \kms (red-yellow-white) and from $-55$ to $-30$ \kms (blue-purple-black). The pigtail is traced by the green dashed line. \label{fig5}}
\end{figure}

\subsection{Kinematics}
Figure 3 shows velocity channel maps of the $^{12}$CO emission.  The pigtail appears in the velocity range from $\VLSR=-80$ \kms\ to $-30$ \kms.  The velocity channels of $\VLSR=-30$ and $-10$ \kms\ are severely contaminated by foreground gas in the Galactic disk.  The first round appears in lower velocities ($-60$ to $-40$ \kms) and the second round, in higher velocities ($-50$ to $-30$ \kms).  These two rounds are connected with less intense emissions at $\VLSR\sim -40$ \kms, supporting the coherent entity of the pigtail.  The pigtail arises from the $-40$ \kms\ cloud (Fig.1{\it a}), which appears in the lower-left edge of the $\VLSR=-40$ \kms\ map (Fig.3).  This $-40$ \kms\ cloud might belong to Arm II of the 120-pc molecular ring (Sofue 2005), which might consist of clouds on the outermost $x_2$ orbit (Binney et al. 1991).  We see another cloud in the lower-left of maps in velocities of $\sim -110$ \kms (hereafter referred to as "the $-110$ \kms\ cloud").  This $-110$ \kms\ cloud appears to belong to the parallelogram (or the 200-pc expanding molecular ring in Kaifu, Kato, \&\ Iguchi 1972), which is considered to be gas in the innermost $x_1$ orbit (Binney et al. 1991).  

We also notice a small clump at $(l, b)\simeq (-0\arcdeg .69, -0\arcdeg .02)$ that appears in the wide velocity range from $\VLSR=-100$ \kms\ to $-60$ \kms.  This broad-velocity clump appears to connect the first round of the pigtail at $\VLSR\sim -40$ \kms\ to the $-110$ \kms\ cloud (Figure 4).  Its broad velocity width and compact spatial size are similar to those of high-velocity compact clouds found in the CMZ (Oka et al. 1998, 1999, 2001a, 2008).  Hereafter, we refer to this broad-velocity clump as "the high-velocity wing" (or simply "the wing").

Figure 5 illustrates the spatial distributions of the  $-110$ \kms\ cloud and the $-40$ \kms\ cloud using $^{12}$CO data from Oka et al. (1998).  The pigtail appears in the $-40$ \kms\ map.  We see complementary spatial distributions of the $-110$ \kms\ and $-40$ \kms\ clouds.  The pigtail arises from the position near the western ends of these clouds, where the wing is also located.  These spatial and velocity structures might indicate the physical interaction between the $-110$ \kms\ and the $-40$ \kms\ clouds.


\subsection{Line Intensity Ratios}
In table 2, we calculated ratios between line intensities integrated for the pigtail, high-velocity wing, $-110$ \kms\ cloud, and the volume excluding these 3 regions and the foreground disk gas from the entire coverage of the data ($|\VLSR|\leq 200$ \kms). Here, we define the foreground disk gas as $ \left( -55 + 10 l \right) $ $\leq |\VLSR| \leq + 15$ \kms, where $l$ is the Galactic longitude. Spatial areas for the integration are indicated in Figure 3, and the velocity ranges are from $-58$ to $-28$ \kms\ for the pigtail, from $-90$ to $-68$ \kms\ for the wing, and from $-100$ to $-130$ \kms\ for the $-110$ \kms\ cloud.  High $^{12}$CO/$^{13}$CO ratios are found in the pigatil and the wing, suggesting that less opaque gas dominates the CO emission from them.  On the contrary, the higher N$_2$H$^+$/HCN ratio, which reflects the opacity of dense molecular gas, indicates a higher opacity of dense gas in the pigtail.  The HCN/CO and CO {\it J} = 3--2/{\it J} = 1--0 ratios are somewhat higher in the pigtail, indicating that density and/or temperature are slightly enhanced.  Similarly, the CS/CO ratio, which reflects the dense gas fraction, is also higher in the pigtail and the wing.  The SiO/$^{13}$CO ratio, which is thought to be a good probe for interstellar shock (e.g., Ziurys, Friberg, \&\ Irvine 1989), is three times larger in the pigtail and the wing than in the surrounding region.  The higher SiO/$^{13}$CO ratio suggests that interstellar shock participates in the formation of the pigtail.  
				   
\begin{deluxetable}{lcccc}
\tiny
\tablecaption{Ratios between integrated line intensities.\label{tbl-2}}
\tablewidth{0pt}
\tablehead{
\colhead{Ratio} & 
\colhead{Pigtail\tablenotemark{a}} &
\colhead{Wing\tablenotemark{a}} &
\colhead{-110 \kms\tablenotemark{a}} &
\colhead{Excluded\tablenotemark{b}} 
}
		   
\startdata
CO {\it J} = 3--2\tablenotemark{c}/1--0 & $0.73(10)$ & $0.86(12)$ & $0.68(10)$ & $0.48(07)$\\
CO/$^{13}$CO & $9.1(1.1)$ & $15(1.8)$ & $4.3(.5)$ & $11(1.3)$\\
HCN/CO & $0.069(8)$ & $0.073(8)$ & $0.062(7)$ & $0.046(5)$\\
SiO/$^{13}$CO & $0.10(1)$ & $0.069(6)$ & $0.013(1)$ & $0.026(2)$\\
N$_{2}$H$^{+}$/HCN & $0.25(3)$ & $0.05(1)$ & $0.20(2)$ & $0.13(2)$\\
CS/CO & $0.077(9)$ & $0.048(6)$ & $0.059(7)$ & $0.032(4)$
\enddata


\tablecomments{Values in parentheses show one standard deviation of uncertainty in the last digits.}
\tablenotetext{a}{Calculated for the area indicated in Figs.2, 3, and 4.}
\tablenotetext{b}{Excluded means the volume excluding the pigtail, wing, -110 \kms cloud, and the foreground disk gas from the entire volume of the observational area.}

\tablenotetext{c}{The CO {\it J} = 3--2 data obtained using ASTE (Oka et al. 2007) were used.}
\end{deluxetable}

\section{Discussion}

\subsection{Observational Facts}
Here, we summarize the observational facts about the pigtail molecular cloud (Fig.5).  \\
1. It is a conical helix gaseous nebula in the CMZ.  \\
2. It has a coherent entity with a continuous velocity field.  \\
3. It arises from the $-40$ \kms\ cloud, which might belong to the outermost $x_2$ orbit.  \\
4. The $-110$ \kms\ cloud, which might belong to the innermost $x_1$ orbit, partially overlaps with the pigtail.  \\
5. The high-velocity wing connects the pigtail at $\VLSR\sim 40$ \kms\ and the $-110$ \kms\ cloud.  \\
6. Density and/or temperature might be higher in the pigtail and the wing than in the $-40$/$-110$ \kms\ clouds.  \\
7. There is evidence of an interstellar shock having had some influence on the formation of the pigtail and the wing.  \\

The helical shape is most likely associated with a twisted magnetic tube.  Therefore, we propose a formation scenario of the pigtail molecular clouds along with the twisting of the vertical magnetic field that penetrates the CMZ.  

\subsection{Physical Conditions}
Before describing the formation scenario, we estimate its physical conditions in the pigtail.  We estimate the gas kinetic temperature from the $^{12}$CO and $^{13}$CO line temperatures in the pigtail under the assumption of the LTE condition and an isotopic abundance ratio [$^{12}$CO]/[$^{13}$CO] = 24 (Langer \&\ Penzias 1990).  For three representative positions (indicated by cross marks in Fig.2{\it a}), $T_{\rm k}=14, 19$, and $36$ K are obtained.  Using these kinetic temperatures and adopting an abundance ratio [$^{13}$CO]/[H$_2$]$=1\times 10^{-6}$ (Lis \&\ Goldsmith 1989), the mass of the pigtail is calculated from the $^{13}$CO {\it J} = 1--0 integrated intensity to be $(2\mbox{--}6)\times 10^5$ $M_{\sun}$ if we assume that $D_{\rm GC}=8.3$ kpc (Gillessen et al. 2009).  This mass is comparable to that of a typical giant molecular cloud.

According to Large Velocity Gradient (LVG) calculations (Goldreich \& Kwan 1974), $R_{\rm HCN/CO} = 0.063$ and $R_{3\mbox{--}2/1\mbox{--}0} = 0.72$ provide $n({\rm H}_2)=10^{3.5}$ cm$^{-3}$ and $N({\rm H_2})/dV=10^{21.0}$ cm$^{-2}(\kms)^{-1}$ if we assume $T_{\rm k}=30$ K and [HCN]/[CO]$=2\times 10^{-3}$ (Tanaka et al. 2009).  Another estimate of the gas density in the pigtail comes from the CO {\it J} = 1--0 integrated intensity using the CO-H$_{2}$ conversion factor, $X_{\rm CO}\equiv N({\rm H}_2)/I_{\rm CO}$, where $N({\rm H}_2)$ is the H$_2$ column density and $I_{\rm CO}$ is the $^{12}$CO integrated intensity.  According to Nagai (2008) and Sofue (2007), we employed $X_{\rm CO}=1\times 10^{20}$ cm$^{-2}$ (K \kms)$^{-1}$, which is approximately one-third of the standard value (Young \&\ Scoville 1991).  Using a typical value of the $^{12}$CO integrated intensity, $I_{\rm CO}\simeq 200$ K \kms\, and assuming that the depth of the pigtail tube is comparable to the width, $\simeq 0\arcdeg .02$ ($3$ pc), we obtained an average gas density $\langle n({\rm H}_2)\rangle\simeq 10^{3.4}$ cm$^{-3}$.  This density coincides with that estimated by the LVG analysis within a factor of 2.  We adopt the LVG density, $n({\rm H}_2)=10^{3.5}$ cm$^{-3}$, in the following discussions.

\begin{figure*}[t]
\epsscale{1.3}
\plotone{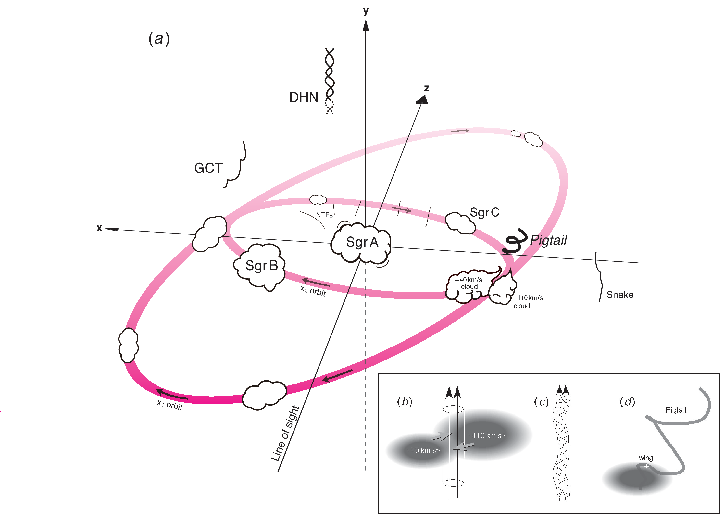}
\caption{Schematic view of the formation of the pigtail molecular cloud. ({\it b}) Vertical tubular magnetic field is sandwiched by clouds in two orbit families. ({\it c}) Two vertical magnetic tubes and trapped gases are gradually twisted. ({\it d}) Heavily twisted tube begins to coil up by the kink instability, forming a helical gaseous nebula.
\label{fig6}}
\end{figure*}

\subsection{Coiled Magnetic Tube Scenario}
\subsubsection{The Scenario}
Here, we hypothesize that the pigtail is associated with a twisted, coiled magnetic tube.  Figure 6 shows the formation process of the pigtail.  At longitudes of $\sim \pm 0.7\arcdeg$, the innermost $x_1$ orbit touches the outermost $x_2$ orbit (Fig.6{\it a}).  Suppose that two clouds on the different orbits sandwich a tube of vertical magnetic field anchored to the Galactic halo at rest.  At the interface between these clouds with different velocities, the velocity shear causes clockwise (viewed from Galactic north) vortices (Fig.6{\it b}).  The vertical magnetic tube is twisted and squeezed by such a vortex, as a result of which it becomes thinner (Fig.6{\it c}).  Furthermore, it begins to coil up as the twist becomes more significant.  This effect is known as the kink instability.  If the energy density of the magnetic field is comparable to or greater than the turbulent energy density of the incident gas cloud, the cloud would become squeezed and coiled together with the twisted magnetic tube.  Because the top and bottom of the cloud have free boundaries for expansion along the field lines, the cloud expands along the coiled magnetic tube, forming a helical gaseous nebula slightly above (below) the midplane (Fig.6{\it d}).  


The above scenario is similar to the "magnetic squeezing mechanism" that was proposed as the formation mechanism of the GCT (Sofue 2007).  He considered the epicyclic rotation of a cloud as the twisting mechanism of the magnetic field.  In our scenario, the field twisting is caused by the velocity shear between the $x_1$ and the $x_2$ orbits.  Thus, the spin velocity of the magnetic tube should be much larger than that of the GCT.  This means that the growth time of the pigtail might be much shorter than that of the GCT.  

\subsubsection{Growth Time Scale of the Pigtail}
Because the velocity field in the pigtail may reflect the motion of gas along the field lines, it cannot move across the field while the gas is trapped by the magnetic tube.  As the energy density of the magnetic field might be comparable to the turbulent energy density of the molecular gas, we estimate the equipartition magnetic field, $B_{\rm eq}=600$ $\mu$G, as a rough lower limit to the field strength in the pigtail.  The equipartition magnetic field and gas density obtained above give an Alfv\'en velocity of $V_{\rm A}\gtrsim 17$ \kms.  We expect that the helical pattern travels with the Alfv\'en velocity.  Because the height of the pigtail is ~30 pc, the growth time scale of the pigtail can be estimated to be $\lesssim 1.8$ Myr.  This is much shorter than the rotational period of the $120$ pc molecular ring, $\simeq 6$ Myr.  Alternatively, a rough estimate of the interaction time between the $-40/-110$ \kms\ clouds gives $40\,\mbox{pc}/70\,\kms \sim 1$ Myr.  If the pigtail has been developed in this timescale, the Alfv\'en velocity is calculated to be $V_{\rm A}\sim 35$ \kms, which gives the magnetic field strength of $B_{\rm eq}\sim 1.2$ mG.  This field is comparable to those that are estimated for nonthermal radio filaments (e.g., Morris \&\ Serabyn 2006).  In either case, as expected, the growth time scale of the pigtail is much shorter than that of the GCT, 5 Myr (Sofue 2007).  

\subsubsection{Support for the Scenario}
The enhancement of the SiO/$^{13}$CO ratio in the pigtail could be attributed to the shock caused by the interaction between molecular materials in different orbits.  The high-velocity wing found in the pigtail supports this idea.  This wing emission is evidence of the physical contact between the $-110$ \kms\ cloud and the $-40$ \kms\ cloud, and therefore of the $x_1/x_2$ intersection.  According to the scenario, we can expect a positive longitude counterpart of the pigtail.  In the opposite longitude, $l\sim +0.7\arcdeg$, we cannot find any helical nebulae above or below the midplane.  Instead, we see a signature of violent collisions between giant molecular clouds (Hasegawa et al. 1994).  In the course of this collision, dense and massive cores may have formed at the interface between the colliding clouds, and their collapse resulted in the current burst of massive star formation in the Sgr B2 complex.  Such an active zone may not be able to develop a well-ordered helix of magnetic tubes.  We can also expect a spinning motion in the rope of the pigtail.  It is not detected in our data sets, however.  This non-detection of the spinning motion could be due to the finite angular resolution ($\sim 20\arcsec = 0.83$ pc) or to the over-twisted morphology of magnetic field lines.  

\subsubsection{Implications}
The discovery of the pigtail may provide the third case of a "helical" magnetic field configuration in the Galactic center environment.  While the other two exhibit clear helical patterns over $70$ pc away from the Galactic plane, the pigtail resides very close to the midplane.  This indicates that the vertical magnetic field in the midplane is not necessarily rigid against deformation due to the turbulent motion of gas clouds in the CMZ.  A vertical field of $10\,\mu$G--mG penetrates the CMZ partially, and the toroidal field may be dominant within the dense cloud complexes in the midplane.  A magnetic tube with a field strength less than several hundred $\mu$G would be deformed by the interaction of the turbulent gas.  A head-on collision between a milligauss magnetic tube and a dense cloud may cause a strong shock, which would accelerate electrons to relativistic energies, thereby allowing the magnetic tube to shine brightly via synchrotron radiation.  This may occur at the root of the vertical filaments of the Radio Arc (Oka et al. 2001b; Tsuboi et al. 1997) and possibly at other nonthermal filaments near the midplane as well.  In either case, the interaction between clouds and the vertical magnetic field plays a key role. The origin of the vertical magnetic field in the Galactic center must be clarified by future observations and theoretical studies.

\section{Conclusions}
This paper reports the discovery and study of a new helical gaseous nebula in the Galactic center region called as the "pigtail"' molecular cloud.  The principal results are summarized as follows: \\
1. The pigtail is a conical helix gaseous nebula with a spatial size of $\sim 20\times 20$ pc$^2$ and a mass of $(2\mbox{--}6)\times 10^5$ $M_{\sun}$.  \\
2. It is located at the position where the innermost $x_1$ orbit and the outermost $x_2$ orbit intersect, and it arises from the $-40$ \kms\ cloud that belongs to the latter orbit.  \\ 
3. Line intensity ratios indicate that the pigtail has slightly higher temperature and/or density than other normal clouds in the CMZ.  \\
4. The higher SiO/$^{13}$CO ratio in the pigtail suggests the influence of a shock in its formation process.  \\
5. The high-velocity wing emission connects the pigtail and a cloud at separate velocities. This wing indicates the physical contact between them.  \\
6. We propose the coiled magnetic tube scenario as a formation process of the pigtail.  According to this scenario, the magnetic field strength in the pigatil should be of the order of a few milligauss, and the age of the pigtail should be $\sim$ Myr.  \\

\acknowledgments
We are grateful to the NRO staff for their excellent support in the 45 m observations.  We also thank Dr. M. Machida for the invaluable discussions and suggestions.  The Nobeyama Radio Observatory is a branch of the National Astronomical Observatory of Japan, National Institutes of Natural Sciences.


\end{document}